\documentclass[onecolumn,usegraphicx,usenatbib,useAMS]{mn2e}
\newcommand{\be}{\begin{equation}}
\newcommand{\ba}{\begin{eqnarray}}
\newcommand{\ee}{\end{equation}}
\newcommand{\ea}{\end{eqnarray}}
\newcommand{\del}{\delta}

\newcommand{\etal}{et al.\ }

\begin{document}
\title[Nonlinear clustering of minihalos]
{Nonlinear clustering during the cosmic Dark Ages and its
effect on the 21-cm background from minihalos}
\author[I. T. Iliev et al.]{Ilian~T.~Iliev,$^1$ Evan~Scannapieco,$^1$ 
Hugo~Martel,$^2$ and Paul~R.~Shapiro$^2$\\
$^1$ Osservatorio Astrofisico di Arcetri, Largo Enrico Fermi 5,
50125 Firenze, Italy\\
$^2$ Department of Astronomy, University of Texas, Austin, TX 78712-1083}

\maketitle

\begin{abstract}

Hydrogen atoms inside virialized minihalos (with $T_{\rm
vir}\leq10^4{\rm K}$) generate a radiation background from redshifted
21-cm line emission whose angular fluctuations reflect clustering
during before and during reionization.  We have shown elsewhere that
this emission may be detectable with the planned Low Frequency Array
(LOFAR) and Square Kilometer Array (SKA) in a flat Cold Dark Matter
Universe with a cosmological constant ($\Lambda$CDM). This is a direct
probe of structure during the ``Dark Ages'' at redshifts $z\ga6$ and
down to smaller scales than have previously been constrained.  In our
original calculation, we used a standard approximation known as the
``linear bias'' [e.g. Mo \& White (1996)]. Here we improve upon that
treatment by considering the effect of nonlinear clustering.  To
accomplish this, we develop a new analytical method for calculating
the nonlinear Eulerian bias of halos, which should be useful for other
applications as well. Predictions of this method are compared with the
results of $\Lambda$CDM N-body simulations, showing significantly
better agreement than the standard linear bias approximation. When
applied to the 21-cm background from minihalos, our formalism predicts
fluctuations that differ from our original predictions by up to 30\%
at low frequencies (high-$z$) and small scales.  However, within the
range of frequencies and angular scales at which the signal could be
observable by LOFAR and SKA as currently planned, the differences are
small and our original predictions prove robust. Our results indicate
that while a smaller frequency bandwidth of observation leads to a
higher signal that is more sensitive to nonlinear effects, this effect
is counteracted by the lowered sensitivity of the radio arrays. We
calculate the best frequency bandwidth for these observations to be
$\Delta\nu_{\rm obs}\sim2$ MHz.  Finally we combine our simulations
with our previous calculations of the 21-cm emission from individual
minihalos to construct illustrative radio maps at $z=9$.

\end{abstract}


\section{Introduction}

Despite striking progress over the past decade, cosmologists have
still failed to see into the ``Dark Ages'' of cosmic time.  Although
the details of this epoch between recombination at redshift
$z\sim10^3$ and reionization at $z\ga6$ are crucial for understanding
issues ranging from early structure formation to the process of
reionization itself, no direct observations of any kind have been made
in this redshift range.  Recently, we made the first proposal for such
direct observations (Iliev et al. 2002, from hereafter Paper I) based
on collisional excitation of the hydrogen 21-cm line in the warm,
dense, neutral gas in virialized minihalos (halos with virial
temperature $T_{\rm vir}\leq10^4{\rm K}$), the first baryonic
structures to emerge in the standard CDM universe.  We showed that
collisional excitation is sufficient to increase the spin temperature
of hydrogen atoms inside minihalos above the temperature the cosmic
microwave background (CMB).  Minihalos should generally appear in
emission with respect to the CMB, producing a background ``21-cm
forest'' of redshifted emission lines, well-separated in frequency.
Unlike all previous works \citep{HR79,SR90,SP93,MMR97,SWMB99,TMMR00},
this mechanism does not require sources of Ly$\alpha$ radiation for
``pumping'' the 21-cm line and decoupling it from the CMB.

In Paper I we calculated the 21-cm emission properties of
individual minihalos in detail, along with the total background and its
large-scale fluctuations due to clustering.  Although individual
lines and the overall background are too weak to be readily
detected, the background fluctuations should be measurable
on $\sim 10'-100'$ scales with the currently-planned
radio arrays LOFAR and SKA.  We demonstrated that such observations
can be used to probe the details of reionization as well as measure
the power spectrum of density fluctuations at far smaller scales than
have been constrained previously.

In the current paper we extend these results to smaller scales.  Our
previous calculations showed that the fluctuation signal increases as
the beam size and frequency bandwidth decrease, which corresponds to
sampling the 21-cm emission from minihalos within smaller volumes.  As
these volumes correspond in turn to length scales that are more
nonlinear, nonlinear effects have the greatest impact on the angular
scales and frequency bandwidths at which the signal is the strongest.

This issue is of particular importance as our previous investigations
relied on the standard, simplified nonlinear bias of \cite{MW96}, which 
breaks down at many of the relevant redshifts and length
scales.  For example, an rms density fluctuation in a Cold Dark Matter
universe with a cosmological constant ($\Lambda$CDM)
at $z = 6 (8),$ is strongly nonlinear ($\sigma(M)\geq 1$) for
$M\leq10^9M_\odot$ ($6.2\times10^7M_\odot$), which roughly corresponds
to the region sampled by beam sizes 9'' (3''), for frequency
bandwidths 12 kHz (3 kHz), respectively.  Nonlinear effects can
influence the predicted background fluctuations on even larger scales,
up to few comoving Mpc (corresponding to few hundred kHz frequency
bandwidths and 1-10 arc min beams), even if these scales are still not
strongly nonlinear.  Additionally, rare halos are always more strongly
clustered than the underlying density distribution (i.e. the bias is
$>1$), again bringing nonlinear issues to the fore.

In order to address these important issues we have carried out a
series of high-redshift N-body simulations of small scale structure
formation, which we use to construct 21-cm line radio maps that
illustrate the expected fluctuations in the emission.  As no current
simulations are able to span the full dynamic range relevant to 21-cm
emission, however, we extend our results by developing a new formalism for
calculating the nonlinear Eulerian bias of halos, which is based on
the Lagrangian bias formalism of Scannapieco \& Barkana (2002,
hereafter SB02). We describe this approach in detail in this paper,
verify it by comparing it with the results of our N-body simulations, 
and apply it to to calculate
improved predictions for the fluctuations in the 21-cm emission.
These are then compared to the predictions given in Paper I,
quantifying the impact of nonlinear effects on minihalo emission.

The structure of this work is as follows.  In \S~\ref{simul} we
describe a set of simulations of minihalo emission at high redshift,
and use these to construct simulated maps at small angular scales.
In \S~\ref{bias_calc} we present our improved calculation of the
Eulerian bias and verify it by comparison with the numerical simulations 
presented in \S~\ref{simul}. In \S~\ref{rms_nonlin} we modify the 
calculation of the radiation background from minihalos to incorporate
the contribution due to nonlinear clustering of sources, according to
the formalism described in \S~\ref{bias_calc}, and describe the results 
of our nonlinear formalism.  Conclusions are given in \S~\ref{conlucions}.

\section{Numerical simulations and simulated 21-cm radio maps}
\label{simul}
We simulated the formation of minihalos in three cubic computational
volumes of comoving size 1~Mpc, 0.5~Mpc, and 0.25~Mpc, respectively.
We used a standard Particle-Particle/Particle-Mesh (P$^3$M)
algorithm (Hockney \& Eastwood 1981), with $128^3$ equal-mass particles,
a $256^3$ PM grid, and a softening length of 0.3 grid spacing. Here and 
throughout this paper we consider a flat $\Lambda$CDM model with 
density parameter $\Omega_0=0.3$, cosmological constant $\lambda_0=0.7$, 
Hubble constant 
$H_0=70\,\rm km\,s^{-1}Mpc^{-1}$, baryon density parameter
$\Omega_b=0.02h^{-2}$ (where $h=H_0/100\,\rm km\,s^{-1}Mpc^{-1}$),
and no tilt. 
The initial conditions where generated using the
transfer function of Bardeen et al. (1986) with the normalization of
Bunn \& White (1997) (for details, see Martel \& Matzner 2000, \S2).
All simulations started at redshift $z=50$ and terminated at redshift
$z=9$. To identify minihalos, we used a standard friends-of-friends
algorithm with a linking length equal to 0.25 times the mean
particle spacing.
We rejected halos composed of 20 particles or less.
Table 1 lists
     the comoving size $L_{\rm box}$ of the box, the comoving value of the
     softening length $\eta$, the total mass $M_{\rm tot}$ inside
     the box, and the particle mass $M_{\rm part}$.

Once the halos are identified, we can use their properties and
distribution to compute the corresponding radio maps. We assign to each
halo a 21-cm flux based on its mass and redshift of formation by
modeling the halos as Truncated Isothermal Spheres
\citep{SIR99,IS01}. The 21-cm fluxes from individual minihalos are
obtained by solving the radiative transfer equation self-consistently
through each halo to obtain the line-integrated flux, as described in
Paper I. Let us consider a cylindrical volume with radius
$R=\Delta\theta_{\rm beam}(1+z)D_A(z)/2$, and length $L_{\rm box}$
, where $D_A(z)$ is the angular diameter distance, $\Delta\theta_{\rm
beam}$ is the beam size of the observation. Then the beam-averaged
differential antenna temperature $\overline{\delta T}_b$ is calculated
using equation~6 of Paper I, which can be discretized as follows:
\begin{equation}
\label{dtb3}
\overline{\delta T}_b({\bf x})={c(1+z)^4\over2\nu_0H(z)}
{1\over\pi R^2L_{\rm box}}
\sum_i
(\Delta\nu_{\rm eff})_i(\delta T_{b,\nu_0})_iA_i
e^{-({\bf x}-{\bf x}_i)^2/2R^2}\,,
\end{equation}
where $H(z)$ is the Hubble constant at redshift $z$, $\nu_0$ is the rest-frame
line frequency, $\pi R^2L_{\rm box}$ is the comoving volume of the beam 
$(\Delta\nu_{\rm eff})_i$, $(\delta T_{b,\nu_0})_i$, and  $A_i$ are the 
effective line width, the line-center differential brightness temperature, 
and the geometric cross-section of halo $i$, respectively, and we 
use Gaussian filter in order to ensure that the contribution of a particular minihalo
to a particular pixel varies smoothly with the location of the minihalo.

\begin{table}
\caption{Parameters of the N-body simulations}
\begin{tabular}{llll}
\hline
$L_{\rm box}$ (Mpc)& $\eta$ (kpc) &$M_{\rm tot}(M_\odot)$ &$M_{\rm part}(M_\odot)$\\
\hline\\[-3mm]
1.0 & 1.172 &	$4.079\times10^{10}$ & $1.945\times10^4$\\[2mm]
0.5 & 0.586 &	$5.099\times10^9$    & $2.431\times10^3$\\[2mm]
0.25 & 0.293 &	$6.374\times10^8$    & $3.039\times10^2$\\
\hline
\end{tabular}
\end{table}

\begin{figure}
\begin{center}
\includegraphics[width=10in]{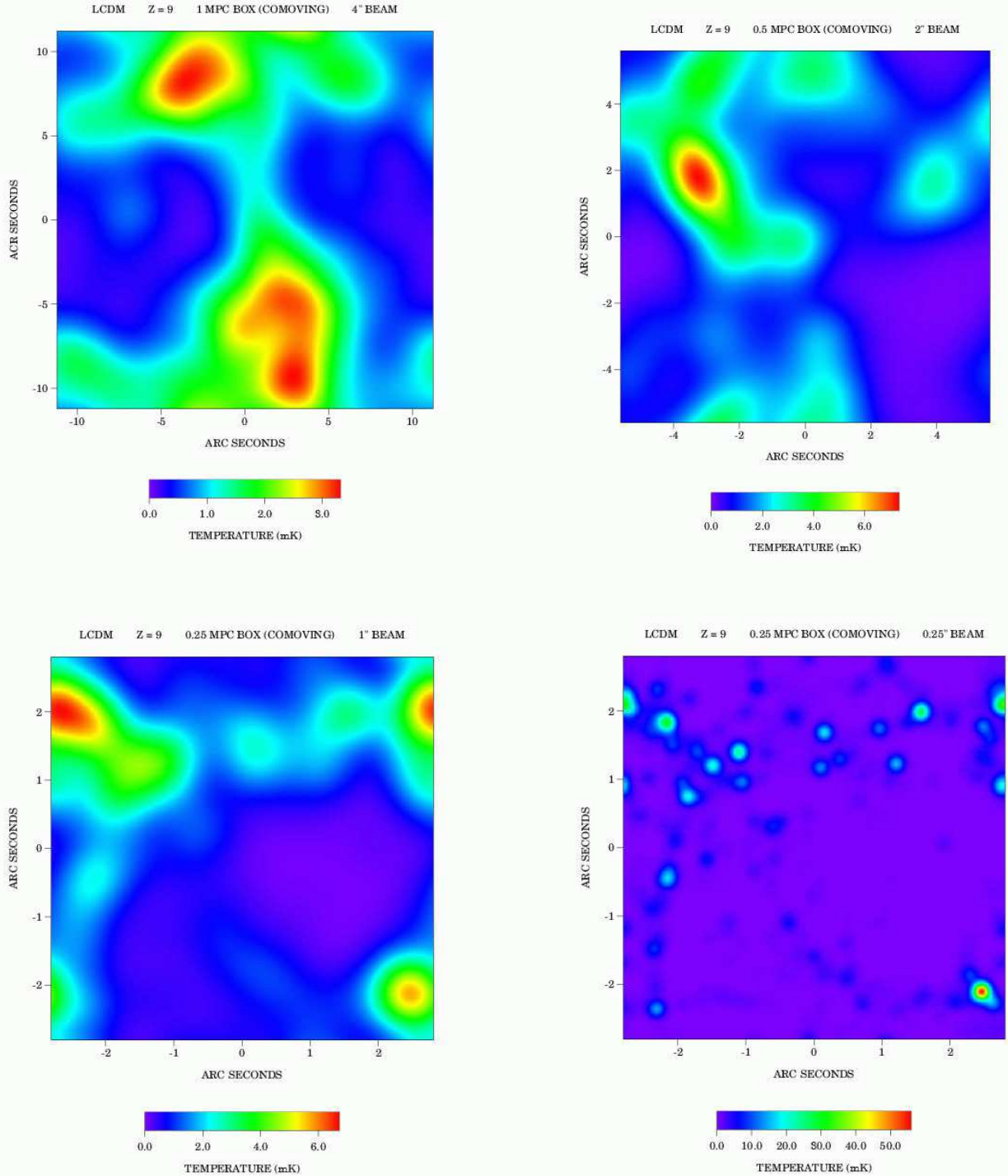}
\end{center}

\caption{
(upper panels) Differential brightness temperature 
radio map of the 21-cm 
emission from minihalos at $z=9$, produced as described in the text from 
$\Lambda$CDM N-body simulation of: (upper left panel) box size 1 Mpc and beam size 
$\Delta\theta_{\rm beam}=4''$, and (upper right panel) box size 0.5 Mpc and beam size 
$\Delta\theta_{\rm beam}=2''.$ 
(lower left panel) box size 0.25 Mpc and $\Delta\theta_{\rm beam}=1''$, and 
(lower right panel) 
box size 0.25 Mpc and $\Delta\theta_{\rm beam}=0.25''$.
\label{tempmap0.25Mpc}
}
\end{figure}

The resulting maps are shown in Figure~\ref{tempmap0.25Mpc}. Due to the small box
sizes that are required to resolve the minihalos, the beams used
to produce the maps are also very small, ranging from 4'' to
0.25''. The fluxes from such small beam sizes are well below the
sensitivity limits of the currently planned radio arrays LOFAR and
SKA. Additionally, the larger-box (1 Mpc and 0.5 Mpc) simulations do
not have sufficient mass resolution to resolve the smallest halos,
while the larger-mass minihalos are not present in the smaller box
simulations. Therefore the flux levels are somewhat underestimated in
all cases, and 
should be considered only as illustrative of fluctuations on very
small scales.

\section{Nonlinear clustering regime and comparison to simulations}
\label{bias_calc}

\subsection{An analytic approach to nonlinear clustering}
\label{bias_analyt}

As our simulations are unable to span a sufficient range of scales to
both resolve all minihalos and reproduce the physical scales
accessible to observations, we adopt instead an alternate technique to
estimate the full nonlinear signal.  We apply an approximate formalism 
that is an extension of the one developed in
SB02. A more complete discussion of this method, refinement of
the basic formalism, and detailed comparisons with simulations will be
presented in a separate paper. In this section we summarize the basic
features of this approach and show that it is sufficient for our
purposes in this investigation.  For an alternative approach see
\citet{CMP98}.

In SB02 the authors extended the Press-Schechter formalism as
reinterpreted by \cite{BC91} to calculate the number density of
collapsed objects in two regions of space initially separated by a
fixed comoving distance.  In this approach virialized halos are
associated with linear density peaks that fall above a critical value
usually taken to be $\delta_{c}=1.686.$ Here the
linear overdensity field is defined as $\delta ({\bf x}) D^{-1}(z)
\equiv \rho({\bf x},z)/\bar\rho(z) - 1$, where $\rho({\bf x})$ is the
linear mass density as a function of position and and $\bar \rho$ is
the mean mass density.  Note that here the linear
overdensity field $\delta({\bf x})$ is described in terms of its value
extrapolated to $z=0$, and its evolution with time is subsumed by the
``linear growth factor'' $D(z) = \delta(0)/\delta{(z)}.$ Other choices
for $\delta_c$ and its evolution have been explored
in the literature, by fitting to simulations
\citep[]{smt01,jenk01} or incorporating the Zel'dovich or
adhesion approximations \citep[]{ls99,men01}.
While these methods improve the accuracy of the Press-Schechter
technique, they provide
no information about nonlinear clustering. Here we strive
to develop a formalism that is applicable to nonlinear clustering
regardless of the recipe chosen for $\delta_{c}$.

Once $\delta_{c}$ determined, the presence of a halo can be associated
with a random walk procedure.  At a given redshift we consider the
smoothed density in a region around a point in space. We begin by
averaging over a large mass scale $M$, or equivalently, by including
only small comoving wavenumbers $k$. We then lower $M,$ adding $k$
modes, and adjusting $\delta(M)$ accordingly.  This amounts to a
random walk in which the ``time variable'' is the variance associated
with the filter mass and the ``spatial variable'' is the overdensity
itself.  This walk continues until
the averaged overdensity crosses $\del_c D^{-1}(z)$ and we assume
that the point belongs to a halo with a mass $M$ corresponding to this
filter scale.

The problem of halo collapse at two positions in space can be
similarly associated with two random walks in the presence of a
barrier of height $\del_c D^{-1}(z)$.  These random walks are
extremely correlated at large scales, and become less and less so
as the modes of smaller and smaller scales are added.  In SB02 this
process was approximated by a completely correlated random walk down
to an overall variance $S$ = $\xi$, where $\xi$ is the correlation
function that accounts for the excess probability of finding a second
halo at a fixed distance from a given halo.  This correlated random
walk was then followed by uncorrelated walks down to the final
variances associated with the masses of the objects, $S=\sigma^2(M_1)$
and $\sigma^2(M_2).$

This procedure lends itself directly to calculating the overall
increase in number density due to infall.  The fact that the
overdensity on large scales will contract the distances between halos
provides an additional contribution at the length scale $x$ such that
$\sigma^2(x)=\xi$, the scale down to which the evolution of the two
points is completely correlated. Let $\delta_0$ be the linear
overdensity at redshift zero at that scale. The probability
distribution of $\delta_0$, $Q_0$ is then simply given by a Gaussian,
with a barrier imposed at $\nu(z) \equiv \delta_c D^{-1}(z)$:
\be
Q_0(\nu, \del_0, \xi) =
[G(\del_0, \xi)-G(2 \nu -  \del_0, \xi)] \,
\theta(\nu - \delta_0),
\label{eq:Qa}
\ee
where $G(\del,\xi)\equiv(2\pi\xi)^{-1/2}e^{-\del^2/2\xi}$, 
and $\theta$ is the Heaviside step function.

We now consider
$f_2(\nu(z),\sigma^2_1,\sigma^2_2,\delta_0,\xi)\, d\sigma^2_1\,
d\sigma^2_2$, the joint probability of having point $A$ in a halo with
mass corresponding to the range $\sigma^2_1$ to $\sigma^2_1+d
\sigma^2_1$ and point $B$ in a halo with mass corresponding to the
range $\sigma^2_2$ to $\sigma^2_2+d \sigma^2_2$, whose random walks
pass though a value of $\delta_0$ at the $\sigma^2(M) = \xi$ scale.
In this case we obtain an expression analogous to equation (37) of SB02
\ba
f_2(\nu(z),\delta_0,\sigma^2_1,\sigma^2_2,\xi(r))
& =& \frac{\partial}{\partial \sigma^2_1} \frac{\partial} {\partial
\sigma^2_2} \left[2 \int_{-\infty}^{\nu(z)} d\del_1 - \int_{-\infty}^{\infty}
d \del_1 \right] 
\left[2 \int_{-\infty}^{\nu(z)} d\del_2 -
\int_{-\infty}^{\infty} d \del_2 \right] 
Q_{12}(\nu,\del_1,\del_2,\sigma^2_1,\sigma^2_2,\xi(r)),
\label{eq:f1} 
\ea
where now $Q_{12} \equiv G(\delta_1 -\delta_0, \sigma^2_1 - \xi)
G(\delta_2 -\delta_0, \sigma^2_2 - \xi)$,
and the reader is referred to SB02 for a more detailed derivation
of $f_2$ from the underlying probability distribution.

Averaging equation (\ref{eq:f1}) over the probability distribution for
$\delta_0$ as given in equation (\ref{eq:Qa}) and
carrying out the partial derivatives we obtain the Eulerian ``bivariate''
number density,
the joint probability of having point $A$ lie in a halo in the mass
range $M_1$ to $M_1 + dM_1$ and point $B$ at a comoving distance
$r$ lie in a halo in the mass range $M_2$ to $M_2 + dM_2$ at a redshift $z$:
\be
\frac{d^2 n^2_{12,E}}{dM_1 dM_2}(r,z) =
\frac{\bar{\rho}}{M_1} \left|\frac{d \sigma^2_1}{d M_1} \right|
\frac{\bar{\rho}}{M_2} \left|\frac{d \sigma^2_2}{d M_2} \right|
\left(1 - \frac{\partial \xi} {\partial \sigma^2_1} \right)
\left(1 - \frac{\partial \xi} {\partial \sigma^2_2} \right)
f_{2,E}(\nu(z), \sigma^2_1,\sigma^2_2,\xi(r)),
\label{eq:2ptm}
\ee
where
\be
f_{2,E}(\nu, \sigma^2_1,\sigma^2_2,\xi) \equiv
\int_{-\infty}^{\nu} d \delta_0 Q_0(\delta_0,\xi) \, g(D^{-1} \delta_0) \,
f(\nu-\delta_0,\xi-\sigma^2_1) \, g(D^{-1} \delta_0) f(\nu-\delta_0,\xi-\sigma^2_2),
\ee
and 
$f(\nu,\sigma^2) \equiv (2\pi)^{-1/2}(\nu/\sigma^3)e^{-\nu^2/2\sigma^2}$.  Here
$g(D^{-1}\delta_0)$ is $\rho/\bar \rho$ at the scale at which the points
$A$ and $B$ are completely correlated.  This can be thought of as the
contraction of a large spherical perturbation that surrounds both
points and contains a total mass $M$ of material, where $\sigma^2(M) =
\xi$.  This contribution is then just $(1+D^{-1}\delta_0)$ in the
linear regime, and for our purposes here we assume $g(D^{-1}\delta_0)
= (1+D^{-1}\delta_0)$ for all values as this reproduces well the 
results from simulations.

Finally we define the Eulerian bias as
\be
b_E^2 \xi \, D^{-2} \equiv \frac{d^2 n^2_{12,E}}{dM_1 dM_2} 
\left( \frac{d n_{1,E}}{dM_1} \frac{d n_{2,E}}{dM_2} \right)^{-1} -1 
\ee
where we divide by $dn_E/dM$, a single point number density that
accounts for the
self correlations between the collapsed peaks and the overdense 
sphere, which are over-counted in equation~(\ref{eq:2ptm}).  
To compute this probability we again carry out an average over the
probability distribution
(\ref{eq:Qa}), but in this case we consider
$f_1(\nu,\delta_0,\sigma^2_1) d\sigma^2_1 =
f(\nu-\delta_0,\sigma^2-\xi) d\sigma^2_1,$
the probability of having a
single point in a halo with a mass corresponding to the range
$\sigma^2_1$ to $\sigma^2_1+d \sigma^2_1$ whose random walk passes
through $\delta_0$ at the $\sigma^2(M) = \xi$ scale.
This gives 
$dn_E/dM=(\bar\rho/M)\,|d\sigma^2\!/dM|\,(1+\partial\xi/\partial\sigma^2)
f_E(\nu,\sigma^2)$
where
\be
f_E(\nu,\sigma^2,\xi) \equiv 
\int_{-\infty}^{\nu} d \delta_0 Q_0(\delta_0,\nu,\xi) 
      	g(\delta_0 D^{-1}) f(\nu -\delta_0,\xi-\sigma^2).
\ee
Combining these expressions yields our final expression for the bias
\be
b_E^2 \xi D^{-2}  = \frac{f_{2,E}(\nu, \sigma^2_1,\sigma^2_2,\xi)}
{f_E(\nu,\sigma^2_1,\xi) \, f_E(\nu_2,\sigma^2_2,\xi)} - 1.
\label{eq:be}
\ee
 
Note that at large distances we can work
to order $\delta_0^2$ to determine the asymptotic
limit as $\xi/\sigma^2 \longrightarrow  0.$  In this limit
\be
g(\delta_0 D^{-1}) f(\nu-\delta_0,\sigma^2-\xi)
\longrightarrow 
\left[1 - 
D^{-1} \, \delta_0 
 \left(\frac{\nu^2}{\delta_c \sigma^2} - \frac{1}{\delta_c} + 1 \right) 
 + {\cal O}(\delta_0^2) \right] f(\nu,\sigma^2-\xi),
\ee  
and the only surviving term in equation (\ref{eq:be}) is the cross
term between terms of order $\delta_0$.  All other terms cancel out
between the numerator and the denominator, giving
\be
b_E^2 = 
\left( 1 + \frac{\nu/\sigma^2_1-1}{\delta_c} \right)
\left( 1 + \frac{\nu/\sigma^2_2-1}{\delta_c} \right),
\label{eq:mowbias}
\ee
Thus our formalism reproduces the usual bias as in \cite{MW96} at large distance, as was
used in Paper I.

Finally, we define the flux-averaged correlation between objects as
\be
\bar b_E^2 \xi D^{-2} = 
\frac{\int dM_1  \, \int  dM_2 \, F(M_1) ({dn}/{dM_1}) \, F(M_2) 
({dn}/{dM_2}) \, b_E^2 \xi D^{-2}}
{\left[\int dM \, F(M) ({dn}/{dM}) \right]^2},
\ee
where $F(M)$ is the line-integrated flux of a minihalo of mass $M$.

\subsection{Nonlinear properties and comparisons with simulations}

\begin{figure}
\includegraphics[width=7in]{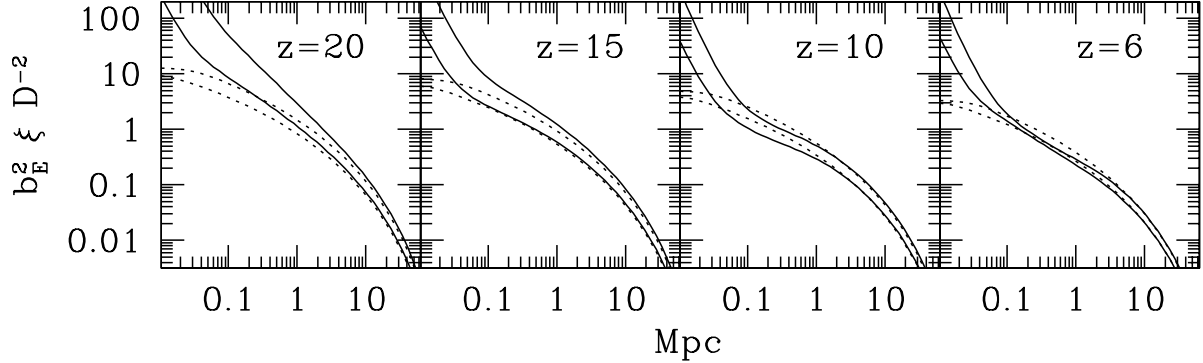}
\caption{Comparison of standard and nonlinear bias.  In all panels,
the solid curves give $b_E^2 D^{-2} \xi$ as calculated in equation~(\ref{eq:be}) 
while the dotted lines give
the standard bias expressions as used in Paper I. For each case, the upper
curves assume a fixed mass of $10^7 M_\odot$, corresponding to a
spherical perturbation with a comoving radius of approximately $38$ kpc,
while the lower
curves assume a fixed mass of $10^6 M_\odot$, corresponding to a comoving radius
of $18$ kpc.}
\label{SB02_vs_MW}
\end{figure}

In Fig.~\ref{SB02_vs_MW} we plot $b_E^2 \xi D^{-2}$ as calculated from
equation~(\ref{eq:be}) and the standard expression,
equation~(\ref{eq:mowbias}), using the fit to the density fluctuation
power spectrum given by \cite{EH99}.  In each panel we show the
behavior of $10^7 M_\odot$ and $10^6 M_\odot$ objects at four
different redshifts, which span the detectable range.  At large
distances, the clustering expressions approach each other
asymptotically, as expected from equation~(\ref{eq:mowbias}). On the
other hand, the solid lines shoot up dramatically at the smallest
distances.  This is because the separation between the halos is
comparable to their radii, and thus the likelihood of finding a second
collapsed halo at the same point becomes infinite as $r
\longrightarrow 0.$ At intermediate distances the behavior is more complex.  For
relatively rare objects, corresponding to high redshifts, the
nonlinear values consistently exceed the standard ones.  For
example, at a distance of 0.1 comoving Mpc, the nonlinear $10^7
M_\odot$ value is almost twice that of the standard result at $z =
20$, while it is only about 70\% of the standard result at $z = 6$.

To compare this behavior to numerical results,
we calculated the correlations between halos for each of our simulations.
In this case $b_E^2 \xi D^{-2}$ is directly comparable to the {\em halo}
correlation function $\xi_{i,j}(r)$, the excess
probability of finding a particle within a mass bin $m_i$ and a
particle within a mass bin $m_j$ separated by a distance $r$.  
This expression is symmetric, so that $\xi_{i,j}(r) = \xi_{j,i}(r)$.
We define $\left<N_{i,j}(r)\right>$ as the mean number of
particles in a mass bin $m_j$,
located within a radius $r$ of a particle in a mass bin $m_i$,
averaged over all particles in a mass bin $M_i$.  
This is is obtained by dividing the total number 
of pairs with separation $r'<r$, by the number $N_i$ of $i$ particles,
\be
\left<N_{i,j}(r) \right>
=\frac{4 \pi N_j}{3 V_{\rm box}} 
	\left[ r^3 + 3 \int_0^r \xi_{i,j}(r') \, r'^2 dr' \, \right],
\label{eq:nij}
\ee
where $N_j$ is the number of
particles of mass $m_j$ and $V_{\rm box}$ is the volume.
Finally, we differentiate equation~(\ref{eq:nij}) to get $\xi_{i,j}(r)$. 
\begin{figure}
\includegraphics[width=4.2in]{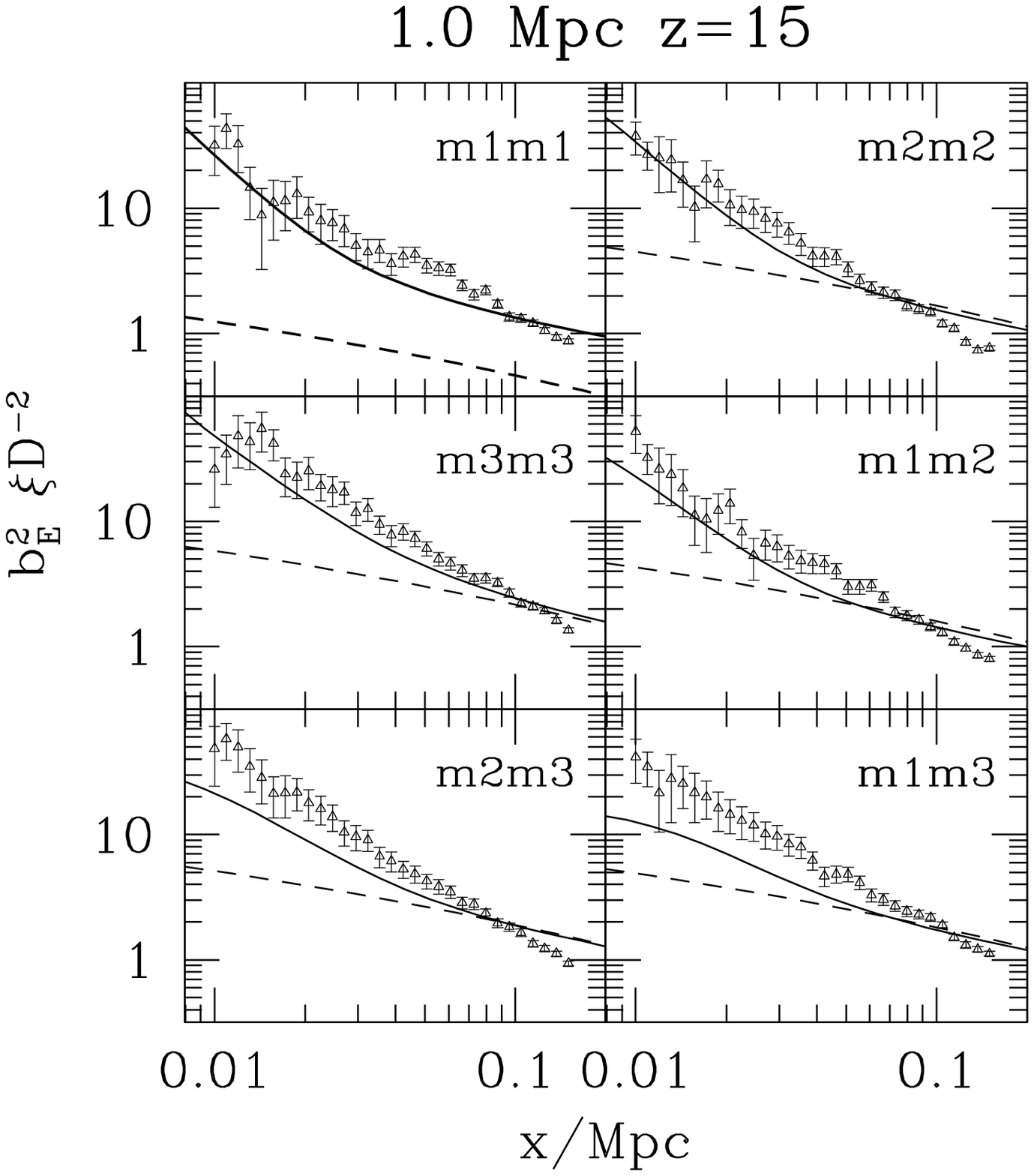}
\hspace{-2cm}
\includegraphics[width=4.2in]{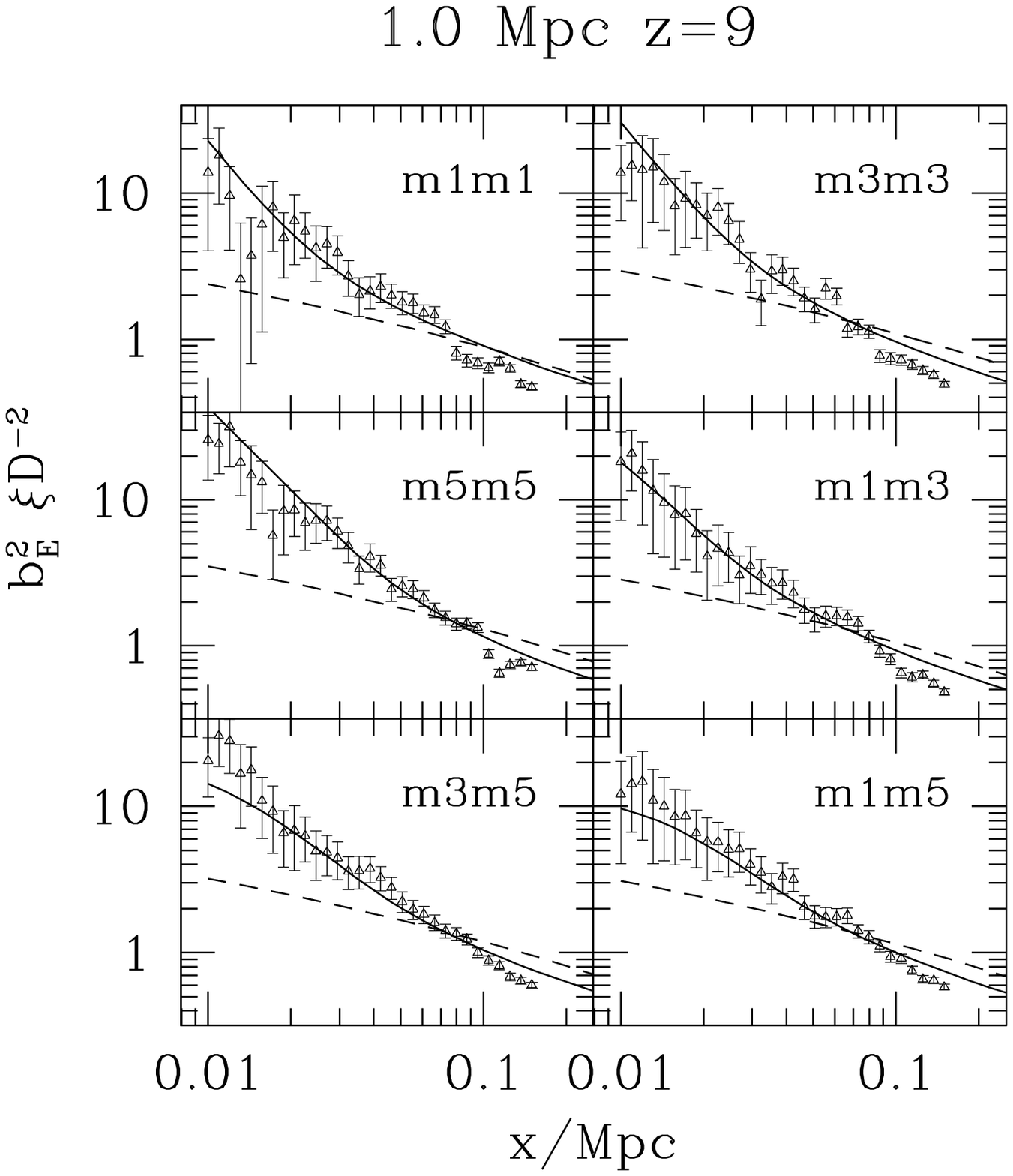}
\caption{ Comparison between nonlinear analytic expressions for the
bias and simulations at two different redshifts.  In the left panels
$z=15$ and the halos are divided into three
mass bins:
   $4.1 \times 10^5 M_\odot \leq m_1 \leq 5.5 \times 10^5 M_\odot$,      
   $5.6 \times 10^5 M_\odot \leq m_2 \leq 1.0 \times 10^6 M_\odot$, and     
   $1.1 \times 10^6 M_\odot \leq m_3 \leq 7.7 \times 10^7 M_\odot$.
In the right panels $z=9$ and halos are divided into five mass
bins, three of which are shown:
   $4.1 \times 10^5 M_\odot \leq m_1 \leq 4.7 \times 10^5 M_\odot$,
   $6.2 \times 10^5 M_\odot \leq m_3 \leq 8.2 \times 10^5 M_\odot$, and 
   $1.3 \times 10^6 M_\odot \leq m_5 \leq 3.4 \times 10^6 M_\odot$.
In both cases, the points are the correlation function as calculated
from the simulations, while the solid lines give our analytic estimates
as  given by equation~(\ref{eq:be}). The error bars are 2-$\sigma$ estimates 
of the statistical noise given by $2 \xi_{i,i}(r)/\sqrt{N_{i,i}(<r)}.$ 
Finally, the dashed lines give correlations as estimated from the standard 
approach, equation~(\ref{eq:mowbias}).}
\label{fig:sim_vs_eul}
\end{figure}

Using these definitions we examined the distribution of halos
in each of our simulations at $z=15$ and $z=9$.  In each case
we divided the halos into mass bins containing roughly the same number ($\sim 600$ )
of objects and compared their correlations to those given by equation~(\ref{eq:be}).  
In the left panels of Fig.~\ref{fig:sim_vs_eul}
we compare the mass averaged correlation between halos,
\be
\bar\xi_{i,j}=\left(\int_{m_i} \frac{dn}{dM_i} \int_{m_j} \frac{dn}{dM_j} b_E^2 
\xi D^{-2} \right)
\left(\int_{m_i} \frac{dn}{dM_i} \int_{m_j} \frac{dn}{dM_j}  \right)^{-1},
\ee
to the numerical value, $\xi_{i,j}$, for objects in three mass bins in
our 1 ${\rm Mpc}^3$ simulation at $z = 15$ (Fig.\ \ref{fig:sim_vs_eul}).
For all mass ranges, there is a good match between the numerical
$\xi_{i,i}$ and our analytic expression, with both values equaling or
exceeding the linear predictions at most distances.  The only case in
which the predictions of the simulations fall below
equation~(\ref{eq:mowbias}), is at large distances, where the
separations between halos begin to approach the simulation box size,
and some damping is to be expected.  Comparisons with our smaller box
simulations yielded comparable results, but with larger damping at
these distances, indicating that the analytic and simulated
correlations are a good match at all reliable separations.  On the
other hand, the cross-correlation functions are a poorer match,
particularly if the mass bins are very different.  For all
correlations however, our nonlinear estimate falls between the
standard and simulated results, indicating that the correlations
between halos exceed those predicted by equation~(\ref{eq:mowbias}),
and that our approach gives a conservative estimate of this excess.

In the right panels of Fig.~\ref{fig:sim_vs_eul}, we repeat this 
comparison with our 1 ${\rm Mpc}^3$ simulation at $z = 9$.
In this case, we divided our halos into five mass bins, 
three of which are shown in this figure. In this
case there is good agreement between the correlations and
cross-correlations obtained from the simulations and our nonlinear
estimates, for all mass scales considered.
In fact the analytical estimates are consistent with the results of
the simulations for all distances that are small relative to the box,
apart from the dip at small distances in the $m_1 m_1$ case, which is
most likely due to small-number statistics.  Note that unlike the
higher redshift case, our nonlinear estimates now fall below the
standard analytic approach at larger distances, as we saw in the lower
redshift cases in Fig.~\ref{SB02_vs_MW}.

\section{Fluctuations of the 21-cm emission from minihalos:  
effects of the nonlinear bias}
\label{rms_nonlin}
The amplitude of $q$-$\sigma$ (i.e. $q$ times the rms value) angular fluctuations
in the differential brightness temperature $\delta T_b$
(or, equivalently, of the flux) in the linear regime are given by 
\begin{equation}
\label{rms_lin}
\langle\delta T_b^2\rangle^{1/2}
	= qb(z)\sigma_p\overline{\delta T_b}
\end{equation}
(Paper I), where $b(z)$ is the mean flux-weighted bias, and 
\begin{eqnarray}
\label{sigma_pk}
\sigma^{2}_{p}=\frac{8D^{-2}(z)}{\pi^2 L^2 R^2}\int^{\infty}_0
 dk \int^1_0 dx \frac{\sin^2 (kLx/2) J^2_1[kR(1-x^2)^{1/2}]}
{x^2(1-x^2)} 
(1+fx^2)^2 \frac{P(k)}{k^2},
\end{eqnarray}
where $P(k)$ is the linear power spectrum at $z=0$,
and the factor $(1+fx^2)^2$, where
$f\approx[\Omega(z)]^{0.6}$ 
is the correction to the cylinder length for the departure from Hubble
expansion due to peculiar velocities \citep{K87}. In order to apply the
formalism developed in \S\ref{bias_calc}, we use that the power spectrum 
is the Fourier transform of the correlation function $\xi(r)$, obtaining
\begin{equation}
\label{sigma_pr}
\sigma^2_p =\frac{32 D^{-2}(z)}{\pi L^2 R^2}
\int^\infty_0 dr r^2 \xi(r) f(r,R,L),
\end{equation}
where
\be
f(r,R,L)\equiv
\int^1_0dx(1+fx^2)^2\int^\infty_0 dk \frac{\sin(kr)}{kr}
\frac{\sin^2({kLx}/{2})}{x^2} \frac{J^2_1[kR(1-x^2)^{1/2}]}
{k^2 (1-x^2)}. 
\ee
Using equations~(\ref{rms_lin}) and (\ref{sigma_pr}) the mean squared 
angular fluctuations become
\be
\langle\delta T_b^2\rangle=\left(\frac{32}{\pi L^2R^2}\right)^2
  \int_0^\infty dr r^2\left[D^{-2}b^2(z)\xi(r)(\overline{\delta T_b})^2\right]
	f(r,R,L).
\ee
We apply the formalism developed in \S\ref{bias_calc} by simply
replacing $\left[D^{-2}b^2(z)\xi(r)(\overline{\delta T_b})^2\right]$ 
with the corresponding quantity
$D^{-2}\overline{b_E^2}\xi(r)(\overline{\delta T_b})^2$ 
for the mean Eulerian flux-weighted
bias calculated in \S\ref{bias_calc}. As we showed, at large
distances the two expressions are equivalent, while at smaller distances
the new formalism better reproduces the numerical results.

\begin{figure}
\includegraphics[width=3in]{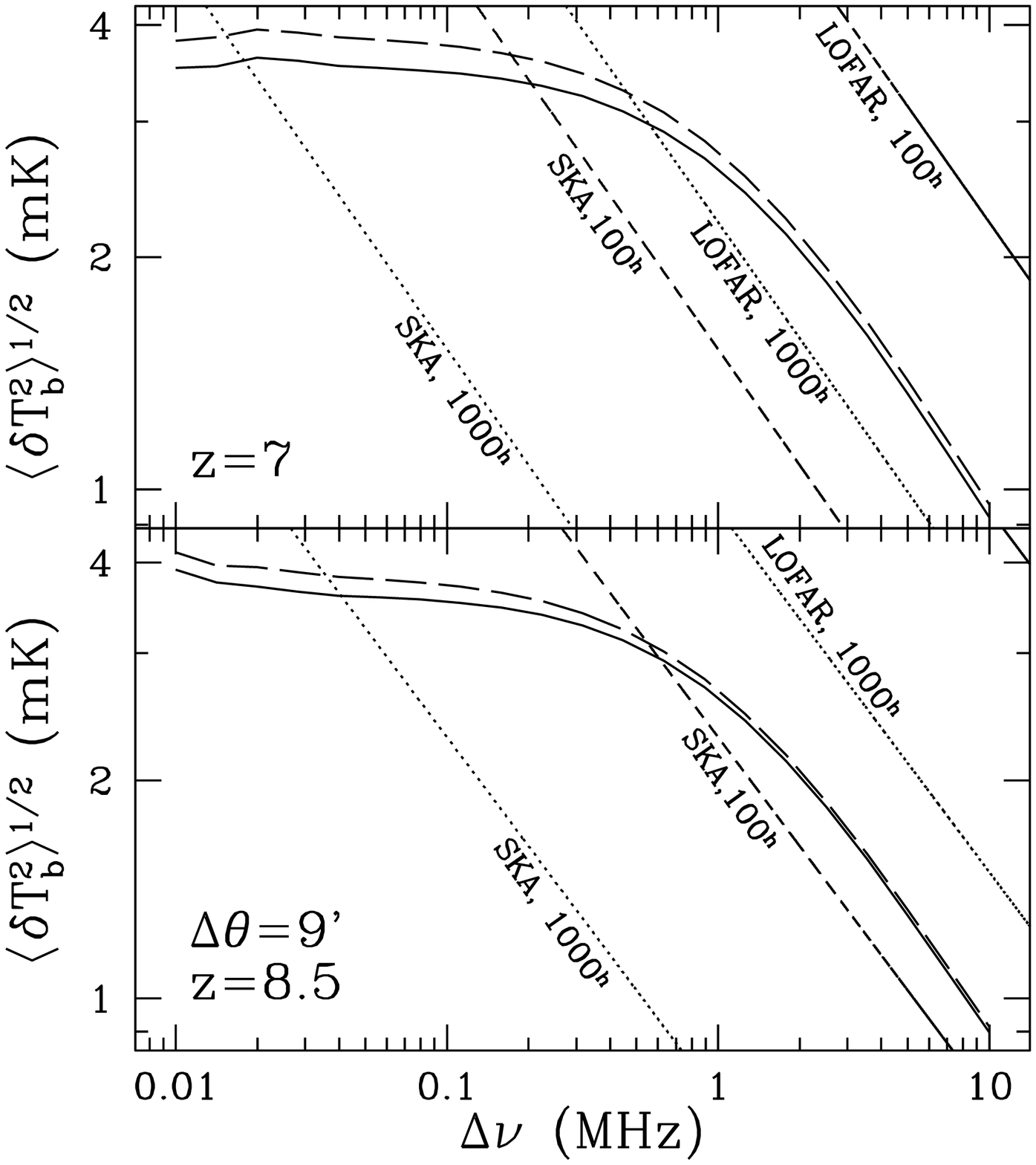}
\includegraphics[width=3in]{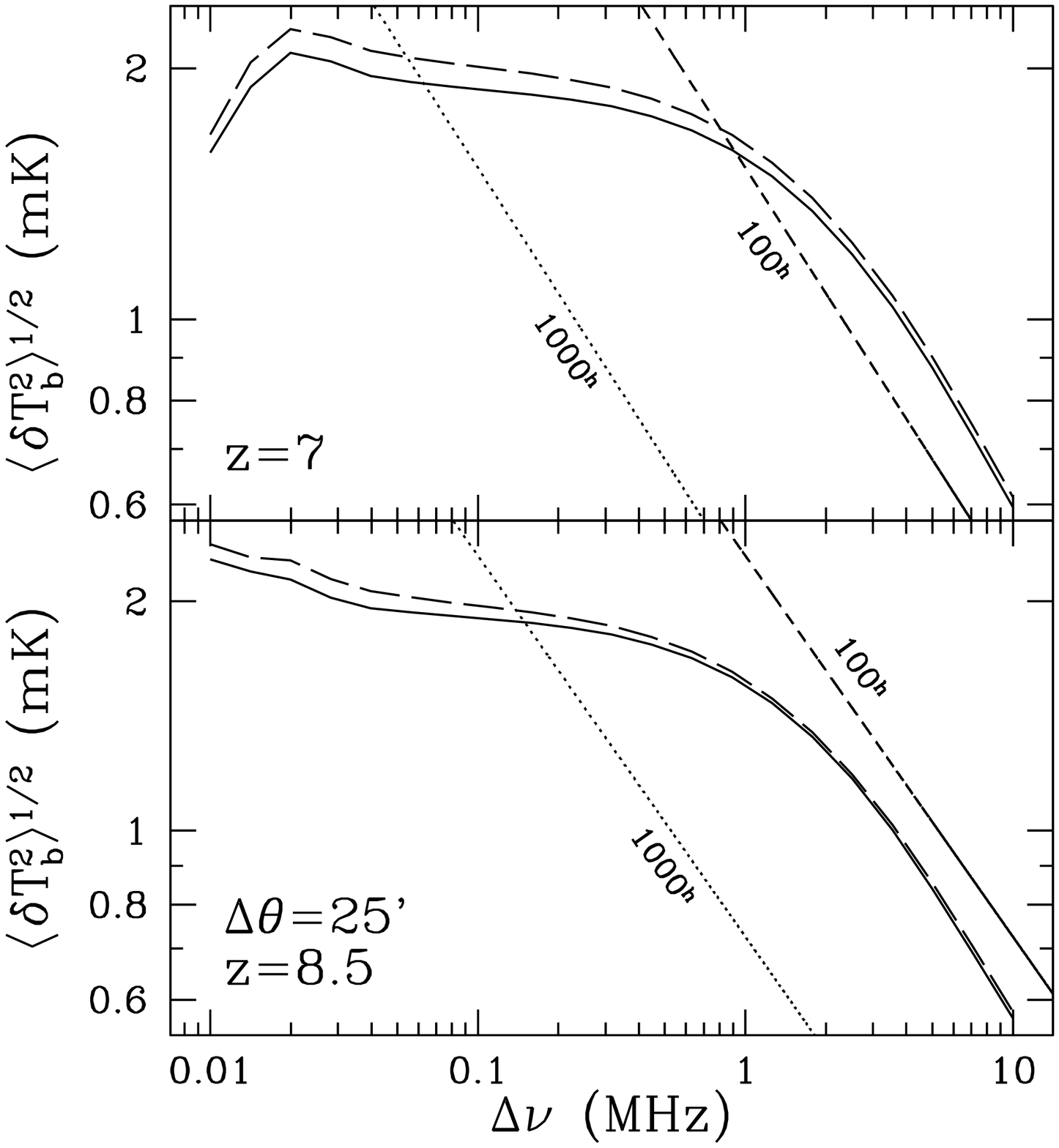}
\caption{Predicted 3-$\sigma$ differential antenna temperature fluctuations
at $z=7$ and $z=8.5$ vs. observer frequency bandwidth
$\Delta\nu_{\rm obs}$ for standard $\Lambda$CDM model using Mo \& White bias 
(long-dashed curves) at angular scales 
(a) (left) $\Delta\theta=9'$, and  (b) (right) $\Delta\theta=25'$.
Also indicated is the predicted sensitivity of LOFAR and SKA 
radio arrays for integration times 100~h 
(dashed) and 1000~h (dotted) (for right panels sensitivity curves for LOFAR and SKA 
are identical) assuming rms sensitivity $\propto \nu^{-2.4}$ 
(see http://www.lofar.org/science).
\label{rms_vs_bandwidth}}
\end{figure}

\begin{figure}
\includegraphics[width=3in]{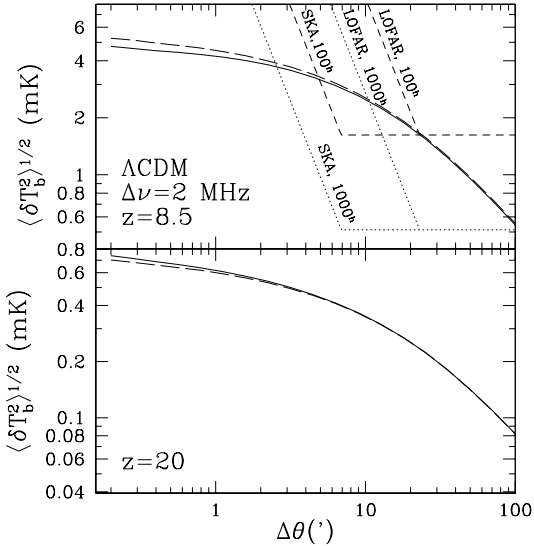}
\caption{Predicted 3-$\sigma$
differential antenna temperature fluctuations
at $z=8.5$ and $z=20$ vs. angular scale $\Delta\theta_{\rm beam}$ 
for standard $\Lambda$CDM model for observer frequency bandwidth
$\Delta\nu_{\rm obs}=2$ MHz 
: using Mo \& White bias (long-dashed curves) and current results (solid curves).
Also indicated is the predicted sensitivity of LOFAR and SKA 
integration times of
100~h (short-dashed lines) and 1000 h (dotted lines), 
with compact sub aperture (horizontal lines) and extended
configuration needed to achieve higher resolution (diagonal lines) 
(see http://www.lofar.org/science and Paper I for details).
\label{rms_vs_theta}}
\end{figure}

\begin{figure}
\includegraphics[width=3in]{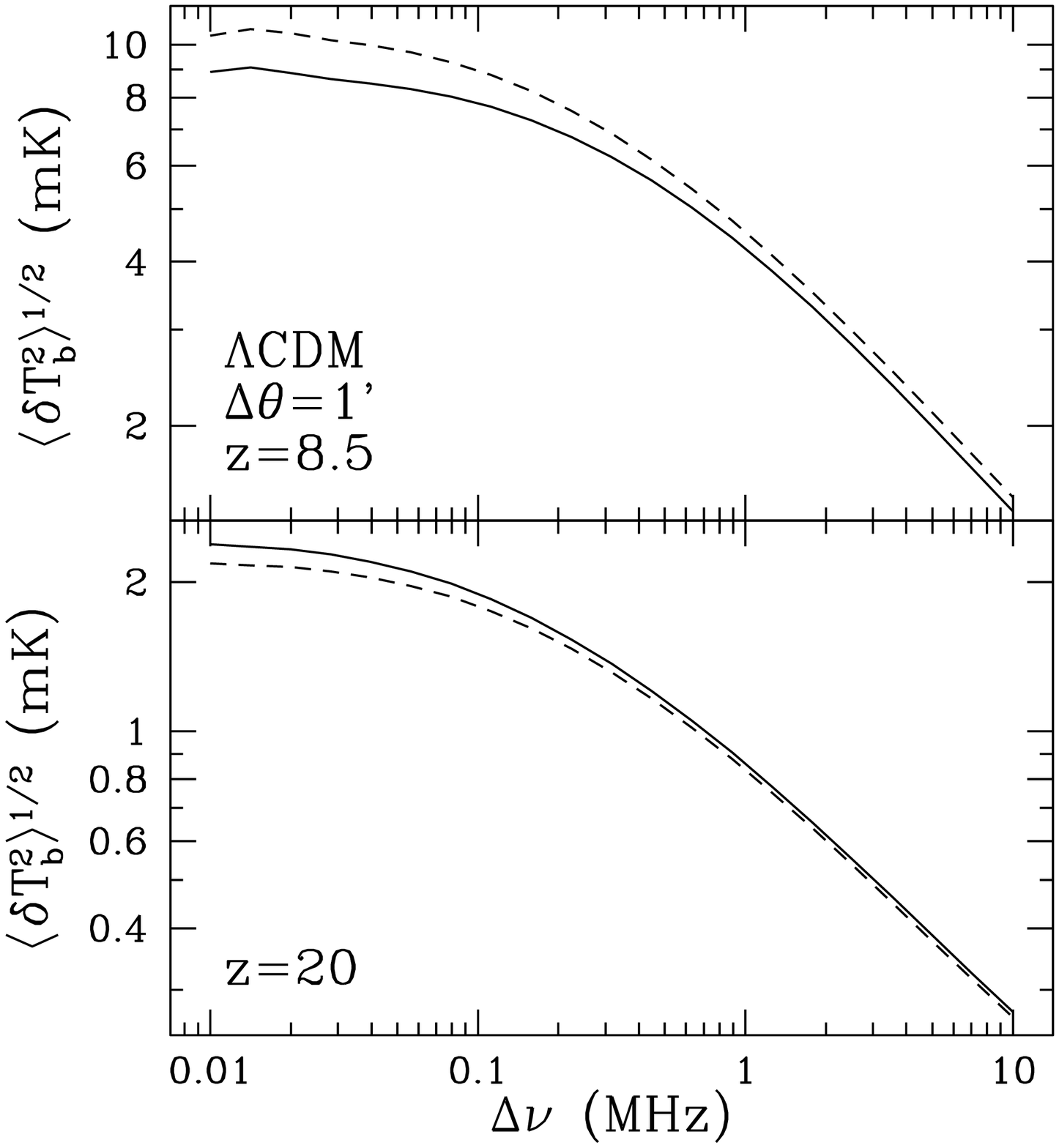}
\includegraphics[width=2.9in]{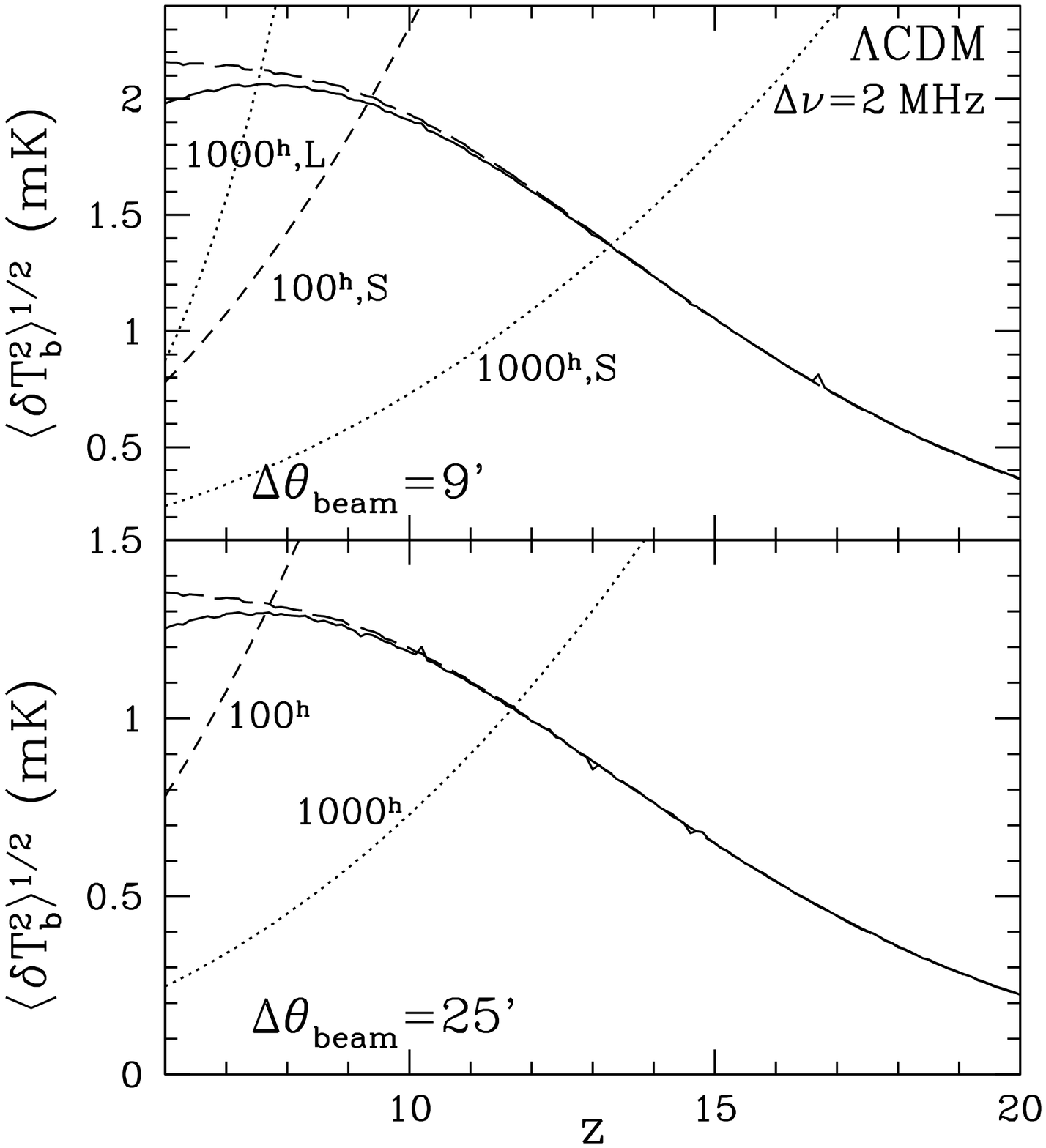}
\caption{(a) (left panels) Predicted 3-$\sigma$
differential antenna temperature fluctuations
at $z=8.5$ and $z=20$ at small angular scales vs. observer frequency bandwidth
$\Delta\nu_{\rm obs}$ for standard $\Lambda$CDM model and for angular scale 
$\Delta\theta=1'$, using Mo \& White bias (long-dashed curves) and current results 
(solid curves). 
(b) (right panels) Predicted 3-$\sigma$ 
differential antenna temperature fluctuations
at $\Delta\theta_{\rm beam}=25'$ vs. redshift $z$ 
for $\Lambda$CDM model, same notation as in a).
The predicted sensitivity  for integration times 100~h 
(dashed) and 1000~h (dotted) of both LOFAR (``L'') and SKA (``S''), are plotted
(for bottom panel, sensitivity curves for LOFAR and SKA are identical)
assuming rms sensitivity 
$\propto \nu^{-2.4}$ (see http://www.lofar.org/science).
\label{rms_vs_deltanu}}
\end{figure}
We start with deriving the optimal frequency bandwidth for observing the
21-cm emission from minihalos. As long as $\Delta\nu_{\rm obs}$ and
the beam size $\Delta \theta_{\rm beam}$ are large enough to provide a
fair sample of the halo distribution, the mean 21-cm flux is
independent of $\Delta\nu_{\rm obs}$, while the fluctuations grow as
$\Delta\nu_{\rm obs}$ decreases. However the sensitivity of the radio 
array deteriorates for smaller bandwidths as 
$\Delta\nu_{\rm obs}^{-1/2}$ (e.g. Shaver et al. 1999; Tozzi et al. 2000).
The results for $3-\sigma$ fluctuations at redshifts $z=7$ and $z=8.5$
vs. observed frequency bandwidth $\Delta\nu_{\rm obs}$
for beam sizes $\Delta\theta_{\rm beam}=9'$ and 25' 
are shown in Fig.~\ref{rms_vs_bandwidth}, along with the results obtained
using the linear bias in order to facilitate comparison. As the bandwidth
increases, the integration time required to detect the signal decreases,
reaching minimum at about $\Delta\nu_{\rm obs}=2$ MHz and remaining roughly
unchanged thereafter. The fluctuations decrease with increasing 
bandwidth, however,hence the optimal bandwidth is about 
$\Delta\nu_{\rm obs}\sim2$ MHz. We see that at large scales the current 
results are largely indistinguishable from the linear results, as expected, 
since at such large scales the bias calculated in \S~\ref{bias_calc} reduces 
to the linear expression in equation~(\ref{eq:mowbias}). For small
values of $\Delta\nu_{\rm obs}$ the differences between the two results 
grow to $\sim 10\%$. However, in the observable range the two results hardly 
differ.

Illustrative results for $3-\sigma$ fluctuations at redshifts $z=8.5$
and $z=20$ vs. the beam size $\Delta\theta_{\rm beam}$ for fixed
observed frequency bandwidth $\Delta\nu_{\rm obs}=2$ MHz
are shown in Fig.~\ref{rms_vs_theta}. The
approximate sensitivities for the LOFAR and SKA radio arrays are also
shown where appropriate (see Paper I for details), but from here on we
concentrate on the differences between the results from the two
approaches and the robustness of our original predictions rather than
observability, which was discussed in Paper~I. At redshifts $z\la15$ using
the linear bias gives an overestimate, 
although a small one compared to the various uncertainties of the
calculation, while at higher redshifts using the linear bias gives a
small underestimate of the fluctuations.

In Fig.~\ref{rms_vs_deltanu} (left panels) we plot the $3-\sigma$ fluctuations at
redshifts $z=8.5$ and $z=20$ vs. the observer frequency bandwidth
$\Delta\nu_{\rm obs}$ for fixed beam size $\Delta\theta_{\rm beam}=1'$. 
We have chosen such a small beam size in order to
investigate the upper limit of the deviation of the results due to the
nonlinear clustering of halos. For small values of both $\Delta\nu_{\rm obs}$ 
and $\Delta\theta_{beam}$ the differences are as large as 20-30\%.
In Fig.~\ref{rms_vs_deltanu} (right panels) we plot the $3-\sigma$ fluctuations for
beam sizes $\Delta\theta_{\rm beam}=9'$ and $25'$ vs. redshift $z$ (i. e. the 
spectrum of fluctuations) for observer frequency bandwidth $\Delta\nu_{\rm obs}=2$ MHz. 
Again, the same patterns emerge, where the linear bias approach overestimates 
the fluctuations at the lower end of the range of considered redshifts by up to 10 \%, 
while underestimating the fluctuations at the higher redshifts, although by a 
smaller fraction.  

\section{Conclusions}
\label{conlucions}

We have developed a new analytical method for estimating the nonlinear
bias of halos and used it to calculate the fluctuations of the 21-cm
emission from the clustering of high-$z$ minihalos.  This method is
likely to be useful in tackling a much wider range of problems that
lie beyond the capabilities of current numerical simulations, and will
be refined and further verified in a future publication.  The
minihalo bias predicted by our method at large scales reproduces the
linear bias, as derived in \cite{MW96}, and is much larger at small
scales, confirming naive expectations.  At intermediate scales,
however, its behavior is more complex and both mass- and
redshift-dependent. For very rare halos at high-$z$ (roughly $z>15$),
the standard linear bias is lower than our nonlinear prediction at all
length scales. At the lower end of the redshift range we consider
(down to $z=6$) the linear bias is higher at intermediate scales (few
hundred kpc to few Mpc comoving) and lower at smaller scales.

We have also compared the predictions of our new method with the results
of N-body numerical simulations, which we used both to verify our
approach and to produce sample radio maps.  Due to the limited
dynamical range of our simulations, however, these maps are only
illustrative of the behavior of the fluctuations on very small scales,
below the sensitivity limits of LOFAR and SKA.  On the scales at which
the simulations are reliable, we find excellent agreement between
these results and our analytical approach.  The most significant
discrepancies occur in the calculation of cross-correlation functions
of very different mass bins at higher redshifts. However, these
departures are relatively modest, and in all cases our method
reproduces the simulation results significantly better than the linear
bias.

Despite these differences, we find our original linear bias
predictions for the fluctuations of the 21-cm emission from minihalos
to be robust at the scales and frequencies corresponding to observable
signals.  The prediction using the flux-weighted nonlinear bias never
departs from the linear prediction by more than few percent in that
range, well within the other uncertainties of the calculation. This
robustness is partly accidental, however, and is due to the nonlinear
bias varying above and below the linear one depending on the length
scale, leading to partial cancellation of the differences when the
correlation function is integrated over the length scales.  For small
observational bandwidths, $\Delta\nu_{\rm obs}$, there is less
cancellation and the differences in the two predictions grow at all
beam sizes, $\Delta\theta_{\rm beam}$. Similarly, if
$\Delta\theta_{\rm beam}$ is small, there is little cancellation, and
the discrepancies are larger at all bandwidths.  In these cases the
linear bias gives an overestimate of the 21-cm emission fluctuations
from minihalos at the low end of the redshift range, and an
underestimate at high-$z$, even at large values of $\Delta\nu_{\rm
obs}$. However, when both the beam size and the frequency bandwidth
are large the differences in the two approaches at small scales are
diluted, and the resulting 21-cm emission fluctuations become
identical.  Finally, we predict that the best observational frequency
bandwidth for improving the chances for detection is $\Delta\nu_{\rm
obs}\sim 2$ MHz.  Thus it may be through a such frequency window that
astronomers get their first glimpses of the cosmological Dark Ages.

\section*{Acknowledgments}
We would like to thank Andrea Ferrara and Rennan Barkana for useful
comments and discussions.  This work was supported in part by the
Research and Training Network ``The Physics of the Intergalactic
Medium" set up by the European Community under the contract
HPRN-CT2000-00126 RG29185, NASA grants NAG5-10825 to PRS and NAG5-10826 to
HM, and Texas Advanced Research Program 3658-0624-1999 to PRS and HM.
ES has been supported in part by an NSF MPS-DRF fellowship.

\end{document}